\documentclass[pre,numerical,reprint,nofootinbib]{revtex4-1}
\usepackage[english]{babel}
\usepackage{microtype}
\usepackage{amsmath,amsthm,amssymb,amsfonts}
\usepackage{mathtools}
\usepackage{siunitx}
\usepackage{tabularx,graphicx}
\usepackage{booktabs}
\usepackage{xr}
\usepackage[breaklinks=true,colorlinks=true,linkcolor=blue,urlcolor=blue,citecolor=blue]{hyperref}

\newcommand{\e}{\mathrm{e}}

\begin{document}

\preprint{}

\title{Efficient generation of random rotation matrices in four dimensions}

\author{Jakob T\'{o}mas Bullerjahn$^\mathparagraph$}
\email{jakob.bullerjahn@biophys.mpg.de}
\affiliation{Department of Theoretical Biophysics, Max Planck Institute of Biophysics, 60438 Frankfurt am Main, Germany}

\author{Bal\'{a}zs F\'{a}bi\'{a}n$^\mathparagraph$}
\email{balazs.fabian@biophys.mpg.de}

\affiliation{Department of Theoretical Biophysics, Max Planck Institute of Biophysics, 60438 Frankfurt am Main, Germany}

\author{Gerhard~Hummer}
\affiliation{Department of Theoretical Biophysics, Max Planck Institute of Biophysics, 60438 Frankfurt am Main, Germany}
\affiliation{Institute of Biophysics, Goethe University Frankfurt, 60438 Frankfurt am Main, Germany}

\date{\today}

\begin{abstract}
Markov-chain Monte Carlo algorithms rely on trial moves that are either rejected or accepted based on certain criteria.  Here, we provide an efficient algorithm to generate random 
rotation matrices in four dimensions (4D) covering an arbitrary pre-defined range of rotation angles.  The matrices can be combined with Monte Carlo methods for the efficient sampling of the SO(4) group of 4D rotations.  The matrices are unbiased and constructed such that repeated rotations result in uniform sampling over SO(4).  4D rotations can be used to optimize the mass partitioning for stable time integration in coarse-grained molecular dynamics simulations and should find further applications in the fields of robotics and computer vision.  
\end{abstract}

\maketitle
\def\thefootnote{$\mathparagraph$}\footnotetext{These authors contributed equally to this work.}\def\thefootnote{\arabic{footnote}}

\section{Introduction}

Four-dimensional (4D) rotations appear in a wide range of problems, from theoretical physics to robotics.  They define the symmetries of two-body motions with inverse-square forces, such as the Kepler problem and the hydrogen atom~\cite{Oliver2004}.  Rotations in 4D have been used to model electromagnetic wave propagation for coherent optical communication~\cite{Karlsson2014}, in algorithms for signal processing~\cite{Borowicz2018}, and to construct equimomental systems of point masses for rigid bodies~\cite{LausSelig2020}, i.e., systems of mass points with preserved total mass, center of mass, and inertia tensor.  They can also be used to formulate various problems in robotics and computer vision, such as hand-eye calibration problems~\cite{WuSun2020} and point-set registration~\cite{SarabandiThomas2022}.  The group SO(4) of 4D rotations can be represented in terms of $4 \times 4$ orthogonal matrices with determinant 1.  A rotation in 4D can be regarded as two distinct rotations in a pair of orthogonal planes~\cite{Lounesto2001} or, equivalently, two orthogonal 3D rotations, because a rotation angle and a normal vector to a plane form an element of SO(3).  This is reflected by the fact that SO(4) is a double cover of $\text{SO(3)} \times \text{SO(3)}$~\cite{Lounesto2001}.  Most 4D rotations are so-called double rotations, where the two rotation angles, $\alpha$ and $\beta$, have distinct values, but isoclinic ($\alpha = \beta$) and simple rotations (either $\alpha = 0$ or $\beta = 0$) can also occur.  

There are multiple ways to generate random 4D rotation matrices. For instance, the Cayley transform
\begin{equation*}
    \mathbf{R}_{4} = (\mathbf{I}_{4} + \mathbf{S}_{4}) (\mathbf{I}_{4} + \mathbf{S}_{4})^{-1}
\end{equation*}
turns any skew-symmetric $4 \times 4$ matrix $\mathbf{S}_{4} = - \mathbf{S}_{4}^{\top}$ into an element of SO(4), where $\top$ indicates the transpose and $\mathbf{I}_{4}$ is the identity matrix.  However, in order to generate uniformly distributed rotation matrices, i.e., which map uniformly onto the 3-sphere, the matrix elements of $\mathbf{S}_{4}$ have to be distributed according to a generalization of the multivariate $t$-distribution~\cite{LeonMasse2006}.  Alternatively, one can make use of the QR decomposition~\cite{Mezzadri2007} or Householder transformations~\cite{Stewart1980} to generate uniformly distributed elements of SO($n$) for $n \geq 2$.  

While the above-mentioned methods can be used to efficiently generate uniformly distributed SO(4) elements, their main drawback is the fact that they do not give practitioners full control over the matrix-generation process.  In particular, none of these methods can be used to exclusively produce small-angle rotations in random orthogonal two-dimensional subspaces.  Instead, numerically expensive decomposition schemes have to be used to explicitly check the size of the rotation angles and then either accept or reject the proposed matrix.  However, small-angle matrices are a prerequisite to perform random walks in 4D rotation space with small steps for efficient Monte Carlo sampling, which are, e.g., needed to search for optimal equimomental systems of certain molecular structures to stabilize the numerics in coarse-grained molecular dynamics simulations~\cite{FabianThallmair2022}.  Possible starting points could be
quaternions, i.e., a 4-tuple describing 3D rotations in terms of an angle and an axis of rotation, because the product of two distinct quaternions results in a 4D rotation matrix~\cite{Lounesto2001, Hanson2006}.  Similarly, it might be tempting to consider Cayley factorization of 4D rotations into a pair of left and right-isoclinic 4D rotation matrices~\cite{Perez-GraciaThomas2017, SarabandiPerez-Gracia2019}.  However, for the resulting rotations to be small, said pairs of quaternions or rotation matrices cannot be chosen arbitrarily, but have to satisfy additional constraints. 

Here, we introduce a geometrically appealing way of constructing rotation matrices for simple, double and isoclinic rotations in 4D with arbitrary rotation angles.  The method exploits the fact that any 4D rotation matrix can be written as the matrix exponential of a skew-symmetric matrix, which can then be decomposed into a sum of two orthogonal skew-symmetric matrices, $\mathbf{A}$ and $\mathbf{B}$, weighted by the rotation angles $\alpha$ and $\beta$~\cite{ErdogduOzdemir2020}.  

The paper is structured as follows: In Sec.~\ref{sec:theory}, the above-mentioned decomposition scheme is briefly reviewed in 3D and 4D, after which we proceed to solve the inverse problem of generating the matrices $\mathbf{A}$ and $\mathbf{B}$.  We also explore the nontrivial distributions of 4D rotation angles and show that $\alpha$ and $\beta$ can essentially be sampled from arbitrary distributions in the small-angle limit.  Section~\ref{sec:algorithm} presents an efficient algorithm (in the sense that it requires only a minimal number of pseudorandom numbers) to construct random small-angle $4 \times 4$ rotation matrices suitable to be used in Markov-chain Monte Carlo sampling schemes.  The algorithm is verified in Sec.~\ref{sec:algorithm-tests} via numerical tests.  A summary of our results can be found in Sec.~\ref{sec:conclusions}, followed by an Appendix that contains various technical details and derivations.

\section{Theory}\label{sec:theory}

In principle, it is possible to construct a random small-angle 4D rotation matrix $\mathbf{R}_{4}$ by combining a double rotation matrix
\begin{equation*}
    \mathbf{R}_{4}'(\alpha, \beta) = \begin{pmatrix}
        \cos(\alpha) & \sin(\alpha) & 0 & 0 \\
        -\sin(\alpha) & \cos(\alpha) & 0 & 0 \\
        0 & 0 & \cos(\beta) & \sin(\beta) \\
        0 & 0 & - \sin(\beta) & \cos(\beta)
    \end{pmatrix}
\end{equation*}
with an arbitrary element $\mathbf{Q}_{4}$ of the orthonormal group O(4) as follows:
\begin{equation}\label{eq:pragmatic-method}
    \mathbf{R}_{4}(\alpha,\beta \ll 1) = \mathbf{Q}_{4} \mathbf{R}_{4}'(\alpha, \beta) \mathbf{Q}_{4}^{\top} \, .  
\end{equation}
While the planes of rotation are well defined for $\mathbf{R}_{4}'$ (the matrix describes rotations in the $xy$ and $zw$ planes), the matrix product in Eq.~\eqref{eq:pragmatic-method} ensures that the planes of rotation for $\mathbf{R}_{4}$ are random.  

The downside of Eq.~\eqref{eq:pragmatic-method} is its computational inefficiency. In a costly process, $\mathbf{Q}_{4}$ could be generated by Gram-Schmidt orthogonalization of four 4D vectors. More efficiently, 
software packages like SciPy~\cite{VirtanenGommers2020} make use of an algorithm, which requires 9 pseudorandom numbers~\cite{Stewart1980} to generate 4D rotation matrices, and Eq.~\eqref{eq:pragmatic-method} requires 2 additional degrees of freedom for the rotation angles.  However, the dimension of SO(4) is 6~\cite{Lounesto2001}, so there is definitely room for further improvement, noting that producing quality random numbers is computationally costly.  

In the remainder of this section, we discuss and develop the theoretical premise for an efficient algorithm to generate small-angle 4D rotations.  In Sec.~\ref{sec:runtime-performance}, we show that our novel algorithm outcompetes Eq.~\eqref{eq:pragmatic-method} in runtime performance, which makes it more viable for applications involving Monte Carlo methods.

\subsection{Matrix decomposition for skew symmetric matrices}

A $\smash{3 \times 3}$ matrix $\mathbf{R}_{3}$, representing 3D rotations about 
the unit vector $\smash{\vec{u} = (-s_{23}, s_{13}, -s_{12})^{\top}}$, $\vert \smash{\vec{u}} \vert = 1$, can be expressed in terms of the exponential of a skew-symmetric matrix
\begin{equation*}
    \mathbf{S}_{3} = 
    \begin{pmatrix}
    0 & s_{12} & s_{13} \\
    -s_{12} & 0 & s_{23} \\
    -s_{13} & -s_{23} & 0
    \end{pmatrix}
\end{equation*}
via the Rodrigues rotation formula
\begin{equation*}
\mathbf{R}_{3}(\gamma) = \e^{\gamma \mathbf{S}_{3}} = \mathbf{I}_{3} + \sin(\gamma) \mathbf{S}_{3} + [1 - \cos(\gamma)] \mathbf{S}_{3}^{2} \, .  
\end{equation*}
Here, $\gamma$ denotes the angle of rotation and $\mathbf{I}_{3}$ is the identity matrix. Due to the Cayley-Hamilton theorem and the constraint on the norm of the vector $\vec{u}$, the matrix $\mathbf{S}_{3}$ satisfies the property $\mathbf{S}_{3}^{3} = - \mathbf{S}_{3}$. 

The eigenvalues of skew-symmetric matrices can either be complex conjugate pairs or zero, hence their rank is always even. Therefore, in 3D, the infinitesimal generator $\mathbf{S}_{3}$ of a rotation has rank 2 and is singular, i.e., $\det(\mathbf{S}_{3}) = 0$. Let $\lambda_{1,2} = \pm i\theta$ and $\lambda_{3} = 0$ be the eigenvalues of $\mathbf{S}_{3}$ with corresponding eigenvectors $\vec{s}_1, \vec{s}_2 \in \mathbb{C}^{3}$ and $\vec{s}_3 \in \mathbb{R}^{3}$, where $\vec{s}_{1}$ and $\vec{s}_{2}$ form a complex conjugate vector pair. The matrix $\mathbf{R}_{3}$ then possesses the same eigenvectors as $\mathbf{S}_{3}$ with eigenvalues $\e^{\lambda_{n}}$ for $n=1,2,3$. The rank-deficiency of $\mathbf{S}_{3}$ expresses the fact that rotations in 3D involve a plane and a unique orthogonal direction along which no rotation takes place, namely the axis of rotation. The non-zero elements of $\mathbf{S}_{3}$ represent rotations in the corresponding planes, e.g., the element $s_{12}$ produces rotation in the $xy$-plane. The plane of rotation can be readily expressed as the span of 
\begin{align*}
    \vec{v}_1 = \frac{1}{\sqrt{2}}  \operatorname{Re}(\vec{s}_1 + \vec{s}_2) \, , & & \vec{v}_2 = \frac{1}{\sqrt{2}}  \operatorname{Im}(\vec{s}_1 - \vec{s}_2) \, ,
\end{align*}
where $ \operatorname{Re}(\cdot)$ and $ \operatorname{Im}(\cdot)$ denote the real and imaginary parts of their arguments, respectively. Setting $\vec{v}_3 = \vec{s}_3$ gives an orthonormal set of vectors in $\mathbb{R}^3$ that satisfy $\vec{v}_i \cdot \vec{v}_j = \delta_{ij}$ for $i,j = 1,2,3$.  

These connections between matrix representations and the geometric meaning of both $\mathbf{S}_{3}$ and $\mathbf{R}_{3}$ also translate to 4D rotations, where any rotation matrix satisfies $\mathbf{R}_{4} = \e^{\mathbf{S}_{4}}$ for a skew-symmetric matrix
\begin{equation*}
    \mathbf{S}_{4} = 
        \begin{pmatrix}
        0 & s_{12} & s_{13} & s_{14} \\
        -s_{12} & 0 & s_{23} & s_{24} \\
        -s_{13} & -s_{23} & 0 & s_{34} \\
        -s_{14} & -s_{24} & -s_{34} & 0
        \end{pmatrix} \, .  
\end{equation*}
The matrix $\mathbf{S}_{4}$ can be uniquely decomposed as~\cite{ErdogduOzdemir2020} 
\begin{equation}\label{eq:S-decomposition}
    \mathbf{S}_{4} = \alpha \mathbf{A} + \beta \mathbf{B}	
\end{equation}
for $\alpha, \beta \geq 0$, $\alpha \neq \beta$, with two rank-2 skew symmetric matrices $\mathbf{A}$ and $\mathbf{B}$, which satisfy
\begin{align}\label{eq:BC-properties}
    \mathbf{A}^{3} = - \mathbf{A} \, , & & \mathbf{B}^{3} = - \mathbf{B} \, , & & \mathbf{A} \mathbf{B} = \mathbf{B} \mathbf{A} = 0 \, .  
\end{align}
For simple rotations, where one of the angles $\alpha, \beta$ vanishes, $\mathbf{S}_{4}$ also has rank 2, but in general it has full rank.  
The corresponding Rodrigues rotation formula reads~\cite{ErdogduOzdemir2020} 
\begin{equation}\label{eq:4D-Rodrigues}
    \begin{aligned}
    \mathbf{R}_{4}(\alpha, \beta) = \e^{\mathbf{S}_{4}}
    &  = \mathbf{I}_{4} + \sin(\alpha) \mathbf{A} + [1 - \cos(\alpha)] \mathbf{A}^{2}
    \\
    & \mathrel{\phantom{=}} + \sin(\beta) \mathbf{B} + [1 - \cos(\beta)] \mathbf{B}^{2} \, .  
    \end{aligned}
\end{equation}
Reference~\onlinecite{ErdogduOzdemir2020} introduces an algorithm to calculate the angles $\alpha$ and $\beta$, as well as the matrices $\mathbf{A}$ and $\mathbf{B}$, from a given skew-symmetric matrix $\mathbf{S}_{4}$.  In what follows, we consider the inverse problem of constructing a pair of matrices $\mathbf{A}$ and $\mathbf{B}$ that satisfy Eq.~\eqref{eq:BC-properties}, which then allows us to use Eq.~\eqref{eq:4D-Rodrigues} to generate random 4D rotation matrices for small-angle rotations.  Our results can be used to efficiently perform a random walk in 4D rotation space, as is required, e.g., for the generation of equimomental systems of particles~\cite{LausSelig2020, FabianThallmair2022}.

\subsection{Constructing the matrix $\mathbf{A}$}\label{sec:constructing-A}

Motivated by the fact that a 4D rotation can be represented by a pair of 3D rotations, we consider two three-dimensional vectors, $\vec{a}_{1}$ and $\vec{a}_{2}$, which should satisfy the relation
\begin{equation*}
    (\vec{a}_{1} \cdot \vec{a}_{2})^{2} = \det(\mathbf{A})
\end{equation*}
for the skew-symmetric matrix
\begin{equation*}
    \mathbf{A} = 
        \begin{pmatrix}
        0 & a_{12} & a_{13} & a_{14} \\
        -a_{12} & 0 & a_{23} & a_{24} \\
        -a_{13} & -a_{23} & 0 & a_{34} \\
        -a_{14} & -a_{24} & -a_{34} & 0
        \end{pmatrix} \, .  
\end{equation*}
There are multiple but equivalent ways of associating the elements of $\vec{a}_{1}$ and $\vec{a}_{2}$ with the $a_{ij}$. Here, we choose $\vec{a}_{1}$ analogous to the vector $\vec{u}$ in the three-dimensional case, which uniquely determines $\vec{a}_{2}$ and we get
\begin{align}\label{eq:u-v-vectors}
    \vec{a}_{1} = 
    \begin{pmatrix}
    -a_{23} \\ a_{13} \\ -a_{12}
    \end{pmatrix} \, , & & \vec{a}_{2} = 
    \begin{pmatrix}
    a_{14} \\ a_{24} \\ a_{34}
    \end{pmatrix} \, .  
\end{align}
Analogous to the 3D case, $a_{ij}$ represents rotation in the $ij$-plane, and the dot product of $\vec{a}_{1}$ and $\vec{a}_{2}$ combines elements representing orthogonal planes.  Furthermore, one can show that $\mathbf{A}^{3} = - \mathbf{A}$ holds whenever $\vec{a}_{1}$ and $\vec{a}_{2}$ satisfy
\begin{align}\label{eq:uv-properties}
    \vec{a}_{1} \cdot \vec{a}_{2} = 0 \, , & & \vec{a}_{1} \cdot \vec{a}_{1} + \vec{a}_{2} \cdot \vec{a}_{2} = 1 \, .  
\end{align}
Requiring $\vec{a}_{1}$ and $\vec{a}_{2}$ to be perpendicular ensures that $\mathbf{A}$ is a rank-2 matrix, i.e., a simple rotation involving a single plane of rotation.  The second requirement of $\vec{a}_{1}$ and $\vec{a}_{2}$ forming a six-dimensional unit vector $\smash{(\vec{a}_{1}^{\top}, \vec{a}_{2}^{\top})^{\top}}$ ensures the decomposition of $\mathbf{S}_{4}$ according to Eq.~\eqref{eq:S-decomposition}, as we shall see in Sec.~\ref{sec:constructing-B}.  

Naively, one could think that the construction of $\mathbf{A}$ requires 6 numbers, i.e., one for every element above the diagonal, but it turns out that only 4 are needed.  As discussed in Appendix~\ref{app:points-on-a-sphere}, this is because we can always choose an arbitrary auxiliary vector $\vec{a}_{1}^{*}$ from the elements of the 2-sphere, which are fully determined by the polar and azimuthal angles.  This accounts for 2 degrees of freedom.  We can then identify two vectors perpendicular to $\vec{a}_{1}^{*}$ that span a plane, in which a second auxiliary vector $\vec{a}_{2}^{*}$ is then constructed with an additional degree of freedom.  Finally, a fourth degree of freedom $R$ serves as the ``mixing ratio'' between $\vec{a}_{1}^{*}$ and $\vec{a}_{2}^{*}$, which results in two vectors $\vec{a}_{1} = \vec{a}_{1}^{*} \sqrt{R}$ and $\vec{a}_{2} = \vec{a}_{2}^{*} \sqrt{1 - R}$ that satisfy the constraints in Eq.~\eqref{eq:uv-properties}, and define a matrix $\mathbf{A}$ according to Eq.~\eqref{eq:u-v-vectors}.

\subsection{Constructing the matrix $\mathbf{B}$ from a known matrix $\mathbf{A}$}\label{sec:constructing-B}

Generating a 4D rotation requires, in general, 6 pieces of information~\cite{Lounesto2001}.  We have already shown that 4 numbers are needed to construct the rank-2 matrix $\mathbf{A}$ of a simple rotation, and the decomposition of Eq.~\eqref{eq:S-decomposition} contains 2 independent angles, so the matrix $\mathbf{B}$ must already be encoded in $\mathbf{A}$.  In fact, the two matrices share the same 4 unique eigenvectors, which, analogous to the 3D case, can be used to construct an orthonormal set of vectors that span two orthogonal planes of rotation.  The matrix $\mathbf{A}$ describes rotations in one plane and $\mathbf{B}$ in the other, and the eigenvalues of $\mathbf{A}$ and $\mathbf{B}$ determine which plane is associated with which matrix.  

In what follows, we shall clarify how the above-mentioned relations between the two matrices can be used to construct $\mathbf{B}$ from a given $\mathbf{A}$, such that their linear combination [Eq.~\eqref{eq:S-decomposition}] gives rise to a rotation matrix according to Eq.~\eqref{eq:4D-Rodrigues}.  To this end, let us consider a skew-symmetric matrix
\begin{equation*}
    \mathbf{B} = 
    \begin{pmatrix}
    0 & b_{12} & b_{13} & b_{14} \\
    -b_{12} & 0 & b_{23} & b_{24} \\
    -b_{13} & -b_{23} & 0 & b_{34} \\
    -b_{14} & -b_{24} & -b_{34} & 0
    \end{pmatrix}
\end{equation*}
that satisfies the property $\mathbf{B}^{3} = - \mathbf{B}$.  According to Sec.~\ref{sec:constructing-A}, there exist two vectors
\begin{align*}
    \vec{b}_{1} = 
    \begin{pmatrix}
    - b_{23} \\ b_{13} \\ - b_{12}
    \end{pmatrix} \, , & & \vec{b}_{2} = 
    \begin{pmatrix}
    b_{14} \\ b_{24} \\ b_{34}
    \end{pmatrix} \, ,
    \end{align*}
for which
\begin{align}\label{eq:pq-properties}
    \vec{b}_{1} \cdot \vec{b}_{2} = 0 \, , & & \vec{b}_{1} \cdot \vec{b}_{1} + \vec{b}_{2} \cdot \vec{b}_{2} = 1 \, ,
\end{align}
must hold for $\mathbf{B}$ to have rank 2.  However, $\vec{b}_{1}$ and $\vec{b}_{2}$ cannot be chosen independently of $\vec{a}_{1}$ and $\vec{a}_{2}$, because of the third condition in Eq.~\eqref{eq:BC-properties}.  Under the constraints of Eqs.~\eqref{eq:uv-properties} and~\eqref{eq:pq-properties}, the resulting equations can be solved to give
\begin{align*}
b_{12} & = \mp a_{34} \, , & b_{13} & = \pm a_{24} \, , & b_{14} & = \mp a_{23} \, , 
\\
b_{23} & = \mp a_{14} \, , & b_{24} & = \pm a_{13} \, , & b_{34} & = \mp a_{12} \, ,
\end{align*}
or, equivalently,
\begin{align*}
\vec{b}_{1} = \pm \vec{a}_{2} \, , & & \vec{b}_{2} = \pm \vec{a}_{1} \, .  
\end{align*}
Note that choosing the top or bottom sign corresponds to choosing between $\mathbf{B}$ and $\mathbf{B}^{\top} = - \mathbf{B}$, which boils down to a positive or negative rotation in the angle $\beta$.  We can therefore, without loss of generality, fix the signs of the elements of $\mathbf{B}$ and allow $\beta$ to take both positive and negative values.

\subsection{Sampling the angles $\alpha$ and $\beta$}\label{sec:sampling-angles}

For a random rotation $\mathbf{R}_{3}(\gamma)$ in 3D, the rotation angle $\gamma \in [0,2\pi)$ is far from being uniformly distributed.  Instead, one has $\gamma \sim [1 - \cos(\gamma)]/2\pi$~\cite{Rummler2002}, where the notation $x \sim p(x)$ implies that $x$ is distributed according to $p(x)$.  In the 4D case, the angles $\alpha, \beta \in [0,2\pi)$ entering Eq.~\eqref{eq:S-decomposition} follow the joint probability distribution function (PDF)~\cite{Rummler2002}
\begin{equation}\label{eq:joint-pdf}
    p(\alpha, \beta) = \frac{[\cos(\alpha) - \cos(\beta)]^{2}}{4 \pi^{2}}
\end{equation}
and are therefore not independent, i.e., $p(\alpha, \beta) \neq p(\alpha) \, p(\beta)$.  However, with the help of the auxiliary variables $u$ and $v$, satisfying $\alpha = u + v$ and $\beta = v - u$, one can uncouple the angles of rotation. This variable substitution decomposes $p(\alpha, \beta)$ into a product of independent and identical distributions for $u, v \in [0, 2\pi]$ with the functional form (see Appendix~\ref{app:sampling-small-angles})
\begin{equation}\label{eq:independent-pdf}
    p(z) = \frac{\sin(z)^{2}}{\pi} \, .  
\end{equation}

In principle, it is possible to expand the corresponding cumulative distribution function (CDF) in the limit of $z \ll 1$, and sample small-angle values for $\alpha$ and $\beta$ using inverse transform sampling, as demonstrated in Appendix~\ref{app:sampling-small-angles}.  However, it turns out that the properties of a large-scale rotation generated by multiple, consecutive, small-angle rotations are mostly unaffected by the small-angle distributions of $\alpha$ and $\beta$.  In fact, for the large-scale rotations to be uniformly distributed over SO(4), one only has to ensure that i)~sufficiently many small-angle rotations are used to cover all of the 3-sphere, and ii)~the small-angle rotations are reversible.  The latter requirement is automatically fulfilled when the vectors $\vec{a}_{1}$ and $\vec{a}_{2}$ are sampled from a uniform distribution on the 2-sphere, because then every 4D-rotation (characterized by two orthogonal planes associated with $\vec{a}_{1}$ and $\vec{a}_{2}$) is as probable as its counter rotation (given by $-\vec{a}_{1}$ and $-\vec{a}_{2}$).  We can therefore choose arbitrary distributions to sample small-angle values for $\alpha$ and $\beta$.

\section{Efficient algorithm for the generation of small-angle 4D rotation matrices}\label{sec:algorithm}

We are now ready to formulate an algorithm to generate random, small-angle, 4D rotation matrices, which can be used in Markov-chain Monte Carlo sampling schemes to explore the space of 4D rotations.  In Sec.~\ref{sec:sampling-angles}, we already mentioned that the distributions, from which small-angle values of $\alpha$ and $\beta$ are drawn from, can be chosen arbitrarily.  Here we will thus make use of uniformly distributed pseudorandom numbers to minimize computational costs.  
One way of generating the random vectors $\vec{a}_{i=1,2}^{*}$ needed to construct the matrices $\mathbf{A}$ and $\mathbf{B}$ would be to apply a random SO(3) element to a pair of orthogonal vectors, e.g., $(1,0,0)^{\top}$ and $(0,1,0)^{\top}$.  However, the efficiency of such a construction scheme would largely depend on the method used to generate the SO(3) elements.  

Instead, we rely on the following algorithm, which is efficient in the sense that it only requires a total of 6 random numbers [corresponding to the dimension of SO(4)]:
\begin{enumerate}
\item{Generate two uniformly distributed random numbers, $R_{1} \sim \mathcal{U}_{[-1,1]}$ and  $R_{2} \sim \mathcal{U}_{[0,2 \pi]}$, where $\mathcal{U}_{[s,t]}$ indicates an independent uniform distribution on the interval $[s,t]$.  The auxiliary unit vector $\vec{a}_{1}^{*}$ is then given by (see Appendix~\ref{app:points-on-a-sphere})
\begin{equation*}
    \vec{a}_{1}^{*} = 
    \begin{pmatrix}
        \sqrt{1 - R_{1}^{2}} \cos(R_{2}) \\
        \sqrt{1 - R_{1}^{2}} \sin(R_{2}) \\
        R_{1} 
    \end{pmatrix} \, .  
\end{equation*}
}
\item{Generate an additional random number $R_{3} \sim \smash{\mathcal{U}_{[0,2 \pi]}}$.  The auxiliary unit vector
\begin{equation*}
    \vec{a}_{2}^{*} = 
    \begin{pmatrix}
        R_{1} \cos( R_{2}) \cos(R_{3}) + \sin( R_{2}) \sin( R_{3}) \\
        R_{1} \sin( R_{2}) \cos(R_{3}) - \cos( R_{2}) \sin( R_{3}) \\
        -\sqrt{1 - R_{1}^{2}} \cos( R_{3})
    \end{pmatrix}
\end{equation*}
is then perpendicular to $\vec{a}_{1}^{*}$ (see Appendix~\ref{app:points-on-a-sphere}). Note that if random number generation is fast compared to the evaluation of trigonometric functions, one can obtain $\sin$ and $\cos$ of $R_{2}$ and $R_{3}$ by drawing two points on a unit circle instead.}

\item{Generate a fourth random number $R_{4} \sim \mathcal{U}_{[0,1]}$, which defines the ``mixing ratio'' between the auxiliary vectors, i.e.,
\begin{align*}
    \vec{a}_{1} = \vec{a}_{1}^{*} \sqrt{R_{4}} \, , & & \vec{a}_{2} = \vec{a}_{2}^{*} \sqrt{1 - R_{4}} \, .  
\end{align*}
}
\item{Generate two additional random numbers, $R_{5}, R_{6} \sim \mathcal{U}_{[0,1]}$, and set
\begin{align*}
    \alpha = \varepsilon R_{5} \, ,
    & & 
    \beta = \varepsilon R_{6} \, ,
\end{align*}
for $\varepsilon \ll 1$.  }
\item{Compute the rotation matrix
\begin{equation*}
    \mathbf{R}_{4} = \e^{\mathbf{S}_{4}} = 
    \begin{pmatrix}
        r_{11} & r_{12} & r_{13} & r_{14} \\
        r_{21} & r_{22} & r_{23} & r_{24} \\
        r_{31} & r_{32} & r_{33} & r_{34} \\
        r_{41} & r_{42} & r_{34} & r_{44} 
    \end{pmatrix}
\end{equation*}
associated with $\mathbf{S}_{4} = \alpha \mathbf{A} + \beta \mathbf{B}$ via the Rodrigues formula [Eq.~\eqref{eq:4D-Rodrigues}].  This corresponds to the following element-wise calculations:
 \begin{align*}
     r_{11} & = ( a_{1,y}^{2} + a_{1,z}^{2} + a_{2,x}^{2} ) \Delta + \cos(\beta) \, ,
     \\
     r_{12} & = h_{12} \Delta - a_{1,z} \sin(\alpha) - a_{2,z} \sin(\beta) \, ,
     \\
     r_{21} & = h_{21} \Delta + a_{1,z} \sin(\alpha) + a_{2,z} \sin(\beta) \, ,
     \\
     r_{22} & = \cos(\alpha) - ( a_{1,y}^2 + a_{2,x}^2 + a_{2,z}^2 ) \Delta
     \\
     r_{13} & = h_{13} \Delta + a_{1,y} \sin(\alpha) + a_{2,y} \sin(\beta) \, ,
     \\
     r_{31} & = h_{31} \Delta - a_{1,y} \sin(\alpha) - a_{2,y} \sin(\beta) \, ,
     \\
     r_{23} & = h_{23} \Delta - a_{1,x} \sin(\alpha) - a_{2,x} \sin(\beta) \, ,
     \\
     r_{32} & = h_{32} \Delta + a_{1,x} \sin(\alpha) + a_{2,x} \sin(\beta) \, ,
     \\
     r_{33} & = \cos(\alpha) - ( a_{1,z}^2 + a_{2,x}^2 + a_{2,y}^2) \Delta
     \\
     r_{14} & = h_{14} \Delta + a_{2,x} \sin(\alpha) + a_{1,x} \sin(\beta) \, ,
     \\
     r_{41} & = h_{41} \Delta - a_{2,x} \sin(\alpha) - a_{1,x} \sin(\beta) \, ,
     \\
     r_{24} & = h_{24} \Delta + a_{2,y} \sin(\alpha) + a_{1,y} \sin(\beta) \, ,
     \\
     r_{42} & = h_{42} \Delta - a_{2,y} \sin(\alpha) - a_{1,y} \sin(\beta) \, ,
     \\
     r_{34} & = h_{34} \Delta + a_{2,z} \sin(\alpha) + a_{1,z} \sin(\beta) \, ,
     \\
     r_{43} & = h_{43} \Delta - a_{2,z} \sin(\alpha) - a_{1,z} \sin(\beta) \, ,
     \\
     r_{44} & = ( a_{2,x}^2 + a_{2,y}^2 + a_{2,z}^2 ) \Delta + \cos(\beta) \, ,
 \end{align*}
where $\vec{a}_{i} = \smash{(a_{i,x}, a_{i,y}, a_{i,z})^{\top}}$, $\Delta = \cos(\alpha) - \cos(\beta)$, and 
\begin{align*}
    h_{12} & = h_{21} = a_{2,x} a_{2,y} - a_{1,x} a_{1,y} \, ,
    \\
    h_{13} & = h_{31} = a_{2,x} a_{2,z} - a_{1,x} a_{1,z} \, ,
    \\
    h_{23} & = h_{32} = a_{2,y} a_{2,z} - a_{1,y} a_{1,z} \, ,
    \\
    h_{14} & = h_{41} = a_{1,z} a_{2,y} - a_{1,y} a_{2,z} \, ,
    \\
    h_{24} & = h_{42} = a_{1,x} a_{2,z} - a_{1,z} a_{2,x} \, ,
    \\
    h_{34} & = h_{43} = a_{1,y} a_{2,x} - a_{1,x} a_{2,y} \, .  
\end{align*}
}
\end{enumerate}
A possible way of speeding up the algorithm is to solely rely on simple rotations.  With $\beta=0$, $\sin(\beta) = 0$, and $\cos(\beta) = 1$, one less random number is required in step 4 and the expressions for $r_{ij}$ in step 5 get simplified.  

The algorithm can also be used to generate uniformly distributed SO(4) elements by replacing the fourth step with (see Appendix~\ref{app:sampling-small-angles})
\begin{enumerate}
\setcounter{enumi}{3}
    \item{Generate two additional random numbers, $R_{5}, R_{6} \sim \mathcal{U}_{[0,1]}$, and find the root of
    \begin{equation*}
        f(z_{i}) = 2 z_{i} - \sin(2 z_{i}) - 8 \pi R_{i}
    \end{equation*}
    for $z_{5} = u$ and $z_{6} = v$.  The angles of rotation are then given by
    \begin{align*}
        \alpha = u + v \, , & & \beta = v - u \, .  
    \end{align*}
    }
\end{enumerate}

\section{Testing the algorithm}\label{sec:algorithm-tests}

It is straightforward to numerically verify that the matrices generated by the algorithm proposed in Sec.~\ref{sec:algorithm} satisfy
\begin{align*}
    \mathbf{R}_{4} \mathbf{R}_{4}^{\top} = \mathbf{I}_{4} \, , & & \det(\mathbf{R}_{4}) = 1 \, ,
\end{align*}
and are therefore elements of SO(4).  In this section, we compare its runtime performance to Eq.~\eqref{eq:pragmatic-method}, and verify that the subsequent application of multiple small-angle rotations gives rise to a uniformly distributed large-scale rotation.

\subsection{Runtime performance}\label{sec:runtime-performance}

The algorithms in Eq.~\eqref{eq:pragmatic-method} and Sec.~\ref{sec:algorithm} were both implemented in Julia~\cite{BezansonEdelman2017} and timed using the \texttt{@benchmark} macro provided by the BenchmarkTools.jl package~\cite{BenchmarkTools}.  To generate the random O(4) elements $\mathbf{Q}_{4}$ entering Eq.~\eqref{eq:pragmatic-method}, we translated a Python implementation~\cite{PythonCode} of an algorithm used to generate SO(4) elements~\cite{Stewart1980} into efficient Julia code.  We tested both small-angle sampling and sampling of the joint angle distribution $p(\alpha, \beta)$ [Eq.~\eqref{eq:joint-pdf}].  The runtime measurements of each algorithm were performed using an installation of Julia v.1.8.0 on a machine with a \SI{1.2}{\giga \hertz} Quad-Core Intel Core i7 processor.  

\begin{table}[h]
\caption{Runtimes of the different algorithms to generate SO(4) rotation matrices in numerical tests.  }
\begin{tabularx}{\linewidth}{X X r}
algorithm & angles & median runtime
\\
\toprule
Sec.~\ref{sec:algorithm} & $\alpha, \beta \sim p(\alpha, \beta)$ & \SI{2.332}{\micro \second}
\\
Sec.~\ref{sec:algorithm} & $\alpha, \beta \sim \smash{\mathcal{U}_{[0,\varepsilon]}}$ &  \SI{207.980}{\nano \second}
\\
Sec.~\ref{sec:algorithm} & $\alpha \sim \smash{\mathcal{U}_{[0,\varepsilon]}}$, $\beta = 0$ &  \SI{182.310}{\nano \second}
\\
Eq.~\eqref{eq:pragmatic-method} & $\alpha, \beta \sim \smash{\mathcal{U}_{[0,\varepsilon]}}$ & \SI{1.618}{\micro \second}
\\
\bottomrule
\end{tabularx}
\label{tab:run_time_comparison}
\end{table}

Our results are collected in Table~\ref{tab:run_time_comparison}.  Using the algorithm of Sec.~\ref{sec:algorithm} to generate small-angle double rotations is an order of magnitude faster than Eq.~\eqref{eq:pragmatic-method}, and only marginally slower than restricting oneself to simple rotations with $\beta = 0$.  

\begin{figure}[t!]
\begin{center}
\includegraphics{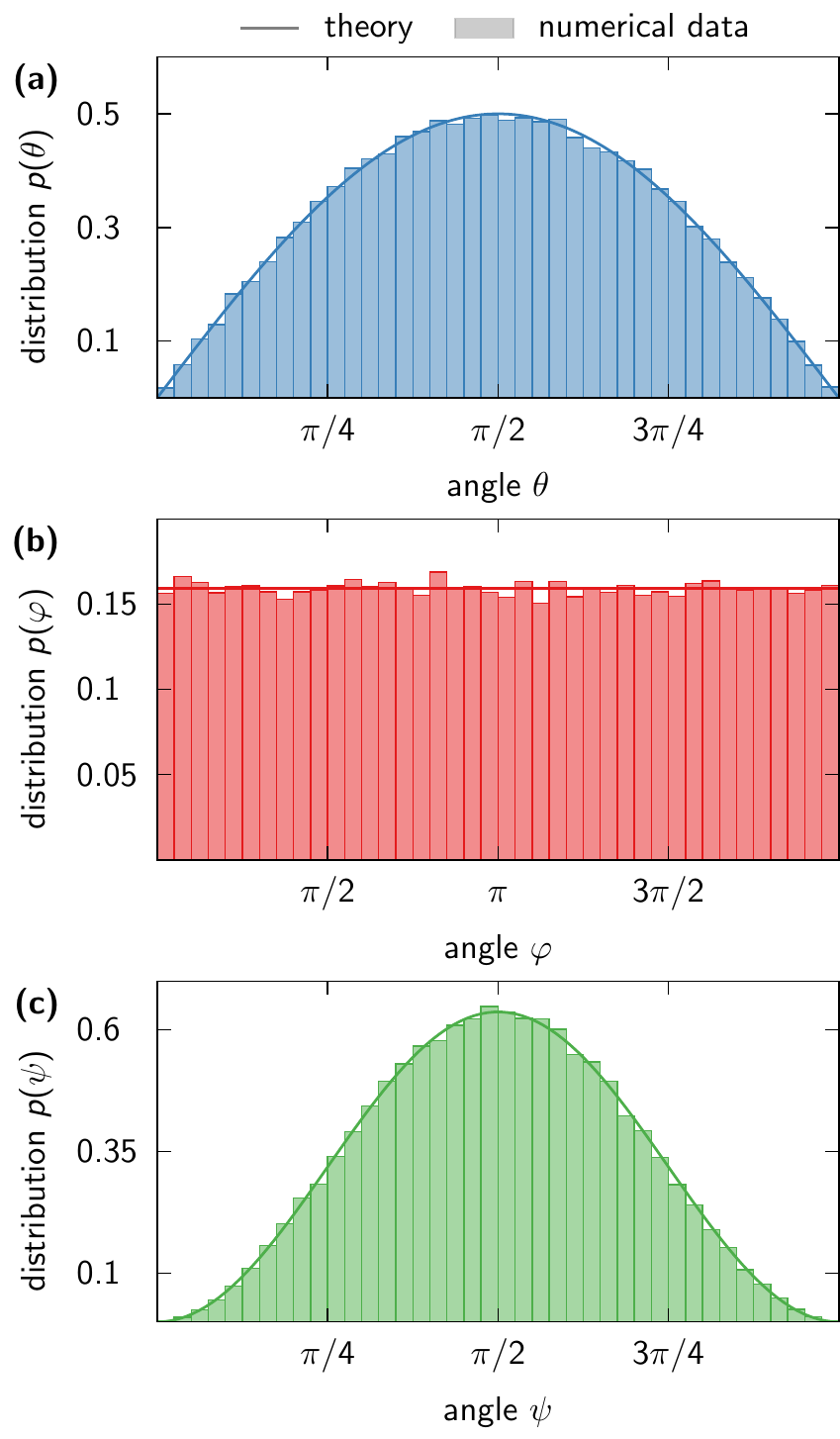}
\caption{Distributions of hyperspherical angles corresponding to a uniform distribution on $S^{3}$.  A set $\smash{\{ \vec{q}'_{m} \}_{m=1}^{M}}$, $M=10^{5}$, of random points on the unit 3-sphere were generated by repeatedly applying $N = 10^{5}$ random, small-angle ($\varepsilon = 0.05$) rotation matrices to a fixed 4-vector $\vec{q} = \smash{(0,0,0,1)^{\top}}$.  The numerical data (histograms) nicely reproduce the angle distributions (a)~$p(\theta) = \sin(\theta)/2$, (b)~$p(\varphi) = 1/2\pi$, and (c)~$p(\psi) = 2 \sin(\psi)^{2} / 2$, which one expects to find if the points $\vec{q}'_{m}$ uniformly cover the surface of the 3-sphere.  }
\label{fig:angle_distributions}
\end{center}
\end{figure}

\begin{figure}[t!]
\begin{center}
\includegraphics{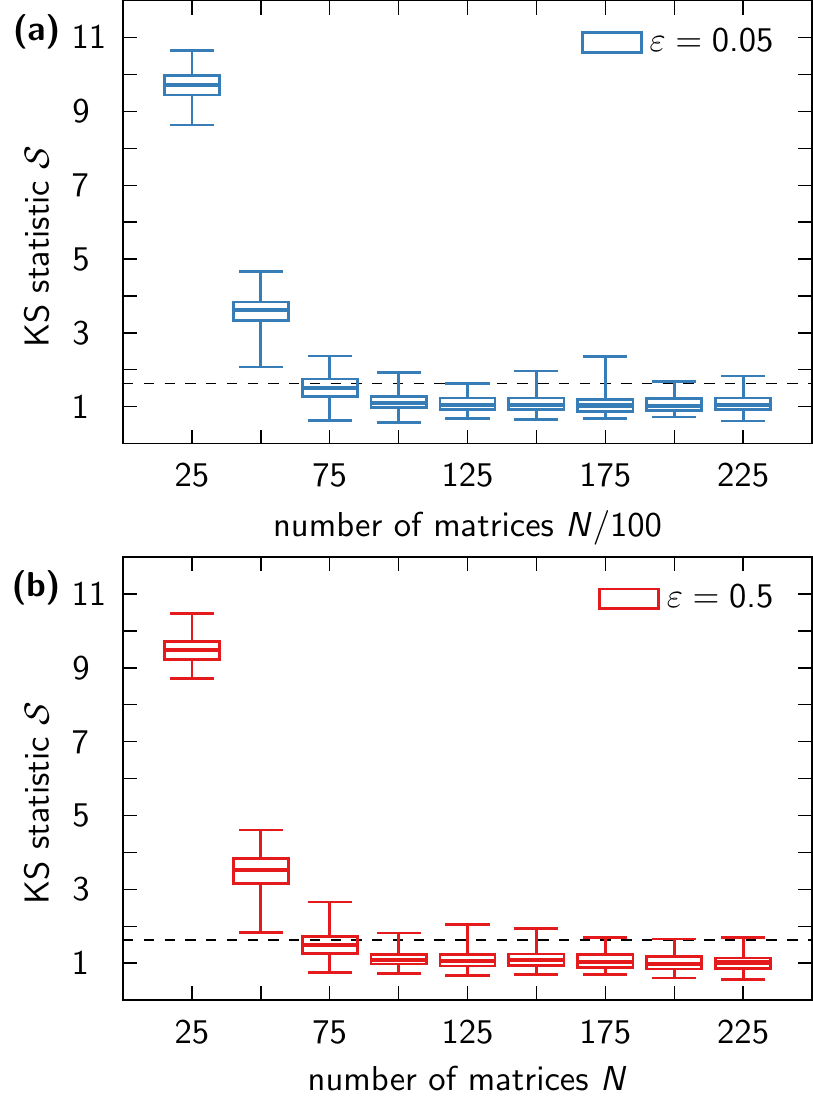}
\caption{Distributions of KS statistics $\mathcal{S} = \max (\mathcal{S}_{\theta}, \mathcal{S}_{\varphi}, \mathcal{S}_{\psi})$ depending on the number of matrices $N$ behind the cumulative rotation.  For each $\mathcal{S}$ a corresponding $p$-value can be calculated and the dashed lines mark $p=0.01$.  We can therefore claim with confidence that the collective rotation of $N$ small-angle rotation matrices generates uniformly distributed SO(4) elements if the corresponding $\mathcal{S}$-distribution mostly lies below the dashed line.  (a)~For $\varepsilon = 0.05$, tens of thousands of small-angle rotation matrices are needed, whereas (b)~$\varepsilon = 0.5$ only requires $N \approx 100$.  }
\label{fig:KS_statistics}
\end{center}
\end{figure}

\subsection{Asymptotic uniformity}\label{sec:testing-uniform-distributions}

If the rotation angles $\alpha$ and $\beta$ are randomly drawn from their joint PDF [Eq.~\eqref{eq:joint-pdf}], we can verify whether the resulting random 4D rotation matrices are uniformly distributed.  This can be realized, e.g., by generating a set of random matrices $\smash{\{ \mathbf{R}_{4}^{(m)} \}_{m=1}^{M}}$ using the algorithm in Sec.~\ref{sec:algorithm} with a modified fourth step, and inspecting the distribution of random points
\begin{equation}\label{eq:large-angle-transform}
    \vec{q}'_{m} = \mathbf{R}_{4}^{(m)} \vec{q}
\end{equation}
that emerges when the matrices are used to rotate a fixed unit 4-vector $\vec{q}$.  For small-valued rotation angles, multiple rotations are needed to compose a large-scale rotation, so Eq.~\eqref{eq:large-angle-transform} has to be replaced with
\begin{equation}\label{eq:small-angle-transform}
    \vec{q}'_{m} = \mathbf{R}_{4}^{(1,m)} \mathbf{R}_{4}^{(2,m)} \dots \mathbf{R}_{4}^{(N,m)} \vec{q} \, .  
\end{equation}
In hyperspherical coordinates, we have (see Appendix~\ref{app:hyperspherical-coordinates}) 
\begin{equation*}
    \vec{q}'_{m} = \begin{pmatrix}
        \sin(\psi) \sin(\theta) \cos(\varphi) \\
        \sin(\psi) \sin(\theta) \sin(\varphi) \\
        \sin(\psi) \cos(\theta) \\
        \cos(\psi)
    \end{pmatrix}
\end{equation*}
because $\vert \vec{q} \vert = 1$, so verifying whether the points $\vec{q}'_{m}$ are uniformly distributed on $S^{3}$ amounts to comparing the distributions of $\{ \theta_{m} \}$, $\{ \varphi_{m} \}$ and $\{ \psi_{m} \}$ to
\begin{align}\label{eq:hyperspherical-coordinate-distributions}
    p(\theta) = \frac{\sin (\theta)}{2} \, , & & p(\varphi) = \frac{1}{2\pi} \, , & & p(\psi) = \frac{2 \sin(\psi)^{2}}{\pi} \, .  
\end{align}

In Fig.~\ref{fig:angle_distributions}, the hyperspherical coordinates of $M=10^{5}$ random points, generated via Eq.~\eqref{eq:small-angle-transform} with $N=10^{5}$ and $\varepsilon=0.05$, are collected into histograms and displayed next to the theoretical predictions of Eq.~\eqref{eq:hyperspherical-coordinate-distributions}.  Visually, the numerical data nicely agree with the analytic PDFs.  The comparison can also be made quantitative with the help of the Kolmogorov–Smirnov (KS) test, which considers the differences
\begin{align*}
    \delta_{+}^{(m)} & = m/M - P(z_{m}) \, ,
    \\
    \delta_{-}^{(m)} & = P(z_{m}) - (m-1)/M \, ,
\end{align*}
between the empirical distribution function and the CDF $P(z)$ of the variable $z \in \{ \theta, \varphi, \psi \}$.  The KS statistic
\begin{equation}\label{eq:KS-statistic}
    \mathcal{S}_{z} = \sqrt{M} \max_{1 \leq m \leq M} \big( \delta_{+}^{(m)}, \delta_{-}^{(m)} \big)
\end{equation}
gives a measure for how likely it is that the sample $\smash{\{ z_{m} \}_{m=1}^{M}}$ was drawn from the PDF $p(z)$.  For the distributions in Fig.~\ref{fig:angle_distributions}, we have
\begin{gather*}
    \begin{aligned}
        P(\theta) = \sin(\theta/2)^{2} \, , & & P(\varphi) = \varphi/2\pi \, , 
    \end{aligned}
    \\
    P(\psi) = [ \psi - \sin(\psi) \cos(\psi) ] / \pi \, ,
\end{gather*}
and the corresponding KS statistics evaluate to $\mathcal{S}_{\theta} \approx 0.90$, $\mathcal{S}_{\varphi} \approx 0.91$ and $\mathcal{S}_{\psi} \approx 1.04$.  The largest KS statistic has an associated $p$-value of $0.23$, so we can be quite confident in our assumption that the cumulative rotation of $N=10^{5}$ small-angle rotation matrices results in a uniformly distributed 4D rotation.  

Figure~\ref{fig:KS_statistics} investigates how many small-angle rotation matrices $N$ are actually needed to construct a uniformly distributed SO(4) element.  We thereby considered 100 sets of points $\smash{\{ \vec{q}'_{m} \}_{m=1}^{M}}$ with $M=1000$, and for each set we evaluated the KS statistic [Eq.~\eqref{eq:KS-statistic}] for the hyperspherical coordinates $\theta$, $\varphi$ and $\psi$.  For $\varepsilon = 0.05$ [Fig.~\ref{fig:KS_statistics}(a)], which was also used to generate the data in Fig.~\ref{fig:angle_distributions}, we find that approximately 10 thousand small-angle matrices will guarantee a uniform distribution, whereas for $\varepsilon = 0.5$ [Fig.~\ref{fig:KS_statistics}(b)] only around 100 matrices are needed.  When used with Monte Carlo methods, the parameter $\varepsilon$ can therefore be seen as a magnitude for the step size, which can be tuned, e.g., to optimize the fraction of accepted moves.  

\section{Conclusions}\label{sec:conclusions}

In this paper, we have presented a geometrically appealing way to construct a 4D rotation matrix from two orthogonal 3D vectors and two angles of rotation (Sec.~\ref{sec:theory}).  The explicit control of the angle variables gives us an advantage over established methods for the generation of SO(4) elements, which we exploit to formulate an algorithm to exclusively generate small-angle rotation matrices in four dimensions (Sec.~\ref{sec:algorithm}).  We also demonstrated (Sec.~\ref{sec:testing-uniform-distributions}) that the repeated application of multiple consecutive small-angle rotations, generated by our proposed algorithm, results in uniform sampling of 4D rotations, as required for an unbiased algorithm. 

The algorithm is efficient in the sense that it only requires the minimal 6 pseudorandom numbers, is an order of magnitude faster than alternative schemes (Sec.~\ref{sec:runtime-performance}), and can be used in combination with Monte Carlo methods to solve optimization problems involving 4D rotations in physics~\cite{LausSelig2020} and engineering.  A practical example is the optimization of the constrained scaffold of a coarse-grained cholesterol model~\cite{FabianThallmair2022}, which is used in combination with the Martini force field~\cite{MarrinkRisselada2007} in molecular dynamics simulations of biological systems.

\begin{acknowledgments}

We thank Prof.~Emeritus Dr.~Andrew J.~Hanson for discussions.  This research was supported by the Max Planck Society. B.F. thanks the Alexander von Humboldt-Foundation for funding.

\end{acknowledgments}

\begin{appendix}

\section{Generating uniformly distributed points on a 3D sphere}\label{app:points-on-a-sphere}

In spherical coordinates only two angles, $\theta \in [0, \pi]$ and $\varphi \in [0,2\pi)$, are needed to define a point on the unit sphere $S^{2} = \smash{ \{ \vec{r} \in \mathbb{R}^{3} \mid \Vert \vec{r} \Vert = 1 \} }$.  The position of the point is given by
\begin{equation}\label{eq:xyz-spherical-coordinates}
\begin{pmatrix}
x \\ y \\ z
\end{pmatrix} = 
\begin{pmatrix}
\sin(\theta) \cos(\varphi) \\
\sin(\theta) \sin(\varphi) \\
\cos(\theta) 
\end{pmatrix}
\end{equation}
in Cartesian coordinates.  Many algorithms exist to generate uniformly distributed points on $S^{2}$. Here, we essentially draw random values for $\theta$ and $\varphi$ from the appropriate distributions.  

A uniform distribution on $S^{2}$ is given in spherical coordinates by $p(\theta,\phi) = \sin(\theta) / 4 \pi$, which implies that the azimuthal angle $\varphi$ is uniformly distributed on the interval $[0,2\pi]$, i.e., $\varphi \sim \mathcal{U}_{[0,2\pi]}$.  The polar angle $\theta$ is distributed according to $p(\theta) = \sin(\theta) / 2$, so in order to draw from this distribution, we employ inverse transform sampling and construct $\theta \sim p(\theta)$ from a uniformly distributed random number $R \in [0,1]$ as follows:
\begin{equation*}
    \theta = \arccos (1 - 2 R) \, .  
\end{equation*}
Note that for $R \sim \mathcal{U}_{[0,1]}$ we have $1 - 2R \sim \mathcal{U}_{[-1,1]}$, so 
\begin{equation}\label{eq:u-vector}
\begin{pmatrix}
x \\ y \\ z
\end{pmatrix} = 
\begin{pmatrix}
\sqrt{1 - R_{1}^{2}} \cos( R_{2}) \\
\sqrt{1 - R_{1}^{2}} \sin( R_{2}) \\
R_{1} 
\end{pmatrix} \, ,
\end{equation}
with $R_{1} \sim \mathcal{U}_{[-1,1]}$ and $R_{2} \sim \mathcal{U}_{[0,2 \pi]}$ is uniformly distributed on $S^{2}$.  Here, we exploited the fact that $\smash{\sin \big( \arccos (x) \big)} = \smash{\sqrt{1 - x^{2}}}$ must hold for $-1 \leq x \leq 1$.  

If Eq.~\eqref{eq:u-vector} is used as a proxy for the auxiliary vector $\smash{\vec{a}_{1}^{*}}$, then there are infinitely many candidates for $\smash{\vec{a}_{2}^{*}}$ lying in a plane spanned by the orthonormal vectors
\begin{align*}
    \begin{pmatrix}
        x' \\ y' \\ z'
    \end{pmatrix} = 
    \begin{pmatrix}
        \cos(\theta) \cos(\varphi) \\
        \cos(\theta) \sin(\varphi) \\
        -\sin(\theta) 
    \end{pmatrix} \, ,
    & & 
    \begin{pmatrix}
        x'' \\ y'' \\ z''
    \end{pmatrix} = 
    \begin{pmatrix}
        \sin(\varphi) \\
        -\cos(\varphi) \\
        0 
    \end{pmatrix} \, .  
\end{align*}
We can randomly choose a candidate of unit length from said plane with a random rotation, giving
\begin{equation*}
    \vec{a}_{2}^{*} = \cos( R_{3}) \begin{pmatrix}
        x' \\ y' \\ z'
    \end{pmatrix} + \sin( R_{3})
    \begin{pmatrix}
        x'' \\ y'' \\ z''
    \end{pmatrix}
\end{equation*}
for $R_{3} \sim \mathcal{U}_{[0,2 \pi]}$.  

\section{Sampling small rotation angles}\label{app:sampling-small-angles}

Equation~\eqref{eq:joint-pdf} in the main text can be rewritten as follows~\cite{Rummler2002}:
\begin{equation*}
    p(\alpha, \beta) = \frac{1}{\pi^{2}} \sin \bigg( \frac{\alpha + \beta}{2} \bigg)^{2} \sin \bigg( \frac{\alpha - \beta}{2} \bigg)^{2} \, ,
\end{equation*}
which invites the coordinate substitutions $u = (\alpha - \beta)/2$ and $v = (\alpha + \beta)/2$ to uncouple the variables.  Both $u$ and $v$ are then distributed according to Eq.~\eqref{eq:independent-pdf} in the main text.  We can expand the diamond-shaped domain of $(u,v) = \smash{\big(} (\alpha - \beta)/2, (\alpha + \beta)/2 \smash{\big)}$, which emerges for $\alpha, \beta \in [0,2\pi]$, to the square $[-\pi, \pi] \times [0,2\pi]$ or, equivalently, $[0, 2\pi] \times [0,2\pi]$.  

The CDF of Eq.~\eqref{eq:independent-pdf} is given by
\begin{equation*}
    P(z) = \frac{2 z - \sin(2 z)}{4 \pi}
\end{equation*}
for $z \in [ 0, 2\pi ]$, and can be used to generate random values for $u$ and $v$ via inverse transform sampling, which amounts to solving
\begin{equation}\label{eq:inverse-transform-sampling}
    P(z) = R_{z}
\end{equation}
for $z \in \{ u,v \}$ and a set of random uniformly distributed numbers $R_{z}=R_{u}, R_{v} \sim \mathcal{U}_{[0,1]}$.  In the small-angle limit, Eq.~\eqref{eq:inverse-transform-sampling} reduces to 
\begin{equation*}
    \varepsilon^{3} R_{z} = \frac{z^{3}}{3 \pi} + \mathcal{O}(z^{4})
\end{equation*}
for $\varepsilon \ll 1$, which gives rise to the following random angles of rotations:
\begin{align*}
    \alpha = \varepsilon ( R_{u}^{1/3} + R_{v}^{1/3} ) \, , & & \beta = \varepsilon ( R_{v}^{1/3} - R_{u}^{1/3} ) \, .  
\end{align*}
However, as discussed in Sec.~\ref{sec:sampling-angles} and demonstrated on explicit examples in Sec.~\ref{sec:algorithm}, the distributions behind $\alpha$ and $\beta$ do not affect the properties of the large-scale rotations resulting from multiple, consecutive, small-angle rotations.

\section{Hyperspherical coordinates}\label{app:hyperspherical-coordinates}

While spherical coordinates are ideal for systems in $\mathbb{R}^{3}$ with rotation symmetries, they can also be generalized to so-called hyperspherical coordinates in higher dimensions.  In 4D, an angle coordinate $\psi \in [0,2\pi]$ is added to the spherical coordinates $r$, $\theta$ and $\varphi$, such that the Cartesian coordinates of a 4-vector are given by
\begin{equation*}
    \begin{pmatrix}
        x \\ y \\ z \\ w
    \end{pmatrix}
    =
    \begin{pmatrix}
        r \sin(\psi) \sin(\theta) \cos(\varphi) \\
        r \sin(\psi) \sin(\theta) \sin(\varphi) \\
        r \sin(\psi) \cos(\theta) \\
        r \cos(\psi)
    \end{pmatrix} \, .  
\end{equation*}
Conversely, one can calculate $r$, $\theta$, $\varphi$ and $\psi$ from a given 4-vector $(x,y,z,w)^{\top}$ as follows:
\begin{align*}
    r & = \sqrt{x^{2} + y^{2} + z^{2} + w^{2}} \, , \\
    \psi & = \arccos (w / r) \, , \\
    \theta & = \arccos (z / \sqrt{r^2 - w^2}) \, , \\
    \varphi & = 
    \begin{cases}
        \arccos(x / \sqrt{r^{2} - w^{2} - z^{2}}) \, , & y \geq 0
        \\
        2\pi - \arccos(x / \sqrt{r^{2} - w^{2} - z^{2}}) \, , & y < 0
    \end{cases} \, .  
\end{align*}
Note that in this paper, we only consider 4-vectors of unit length, so $r \equiv 1$.  

\end{appendix}


\begin{thebibliography}{20}

\bibitem{Oliver2004} D. Oliver, \emph{The Shaggy Steed of Physics}, (Springer-Verlag, New York, 2004).  

\bibitem{Karlsson2014} M. Karlsson, ``Four-dimensional rotations in coherent optical communications'', J. Light. Technol. \textbf{32}, 1246-1257 (2014).  

\bibitem{Borowicz2018} A. Borowicz, ``On using quaternionic rotations for indpendent component analysis'' in \emph{Signal Processing: Algorithms, Architectures, Arrangements, and Applications (SPA 2018)}, 114-119 (IEEE, 2018).  

\bibitem{LausSelig2020} L. P. Laus and J. M. Selig, ``Rigid body dynamics using equimomental systems of point-masses'', Acta Mech. \textbf{231}, 221-236 (2020).  

\bibitem{WuSun2020} J. Wu, Y. Sun, M. Wang, and M. Liu, ``Hand-eye calibration: 4-D procrustes analysis approach'', IEEE Trans. Instrum. Meas. \textbf{69}, 2966-2981 (2020).  

\bibitem{SarabandiThomas2022} S. Sarabandi and F. Thomas, ``Approximating displacements in $\mathbb{R}^{3}$ by rotations in $\mathbb{R}^{4}$ and its application to pointcloud registration'', IEEE Trans Robot. \textbf{38}, 2652-2664 (2022).  

\bibitem{Lounesto2001} P. Lounesto, \emph{Clifford Algebras and Spinors}, (Cambridge University Press, Cambridge, 2001).  

\bibitem{LeonMasse2006} C. A. Le\'{o}n, J.-C. Mass\'{e}, and L.-P. Rivest, ``A statistical model for random rotations'', J. Multivar. Anal. \textbf{97}, 412-430 (2006).  

\bibitem{Mezzadri2007} F. Mezzadri, ``How to generate random matrices from the classical compact groups'', Not. Am. Math. Soc. \textbf{54}, 592-604 (2007).  

\bibitem{Stewart1980} G. W. Stewart, ``The efficient generation of random orthogonal matrices with an application to condition estimators'', SIAM J. Numer. Anal. \textbf{17}, 403-409 (1980).  

\bibitem{FabianThallmair2022} B. F\'{a}bi\'{a}n, S. Thallmair, and G. Hummer, ``Optimal bond-constraint topology for molecular dynamics simulations of cholesterol'', J. Chem. Theory Comput., \url{https://doi.org/10.1021/acs.jctc.2c01032}.  

\bibitem{Hanson2006} A. J. Hanson, \emph{Visualizing quaternions}, (Morgan Kaufmann Publishers, Burlington, MA, 2006).  

\bibitem{Perez-GraciaThomas2017} A. Perez-Gracia and F. Thomas, ``On Cayley’s factorization of 4D rotations and applications'', Adv. Appl. Clifford Algebras \textbf{27}, 523–538 (2017).  

\bibitem{SarabandiPerez-Gracia2019} S. Sarabandi, A. Perez-Gracia, and F. Thomas, ``On Cayley's factorization with an application to the orthonormalization of noisy rotation matrices'', Adv. Appl. Clifford Algebras \textbf{29}, 49 (2019).  

\bibitem{ErdogduOzdemir2020} M. Erdo\v{g}du and M. \"{O}zdemir, ``Simple, double and isoclinic rotations with a viable algorithm'', Math. Sci. Appl. E-Notes \textbf{8}, 11-24 (2020).  

\bibitem{VirtanenGommers2020} P. Virtanen, R. Gommers, T. E. Oliphant, M. Haberland, T. Reddy, D. Cournapeau, E. Burovski, P. Peterson, W. Weckesser, J. Bright, S. J. van der Walt, M. Brett, J. Wilson, K. J. Millman, N. Mayorov, A. R. J. Nelson, E. Jones, R. Kern, E. Larson, C. J. Carey, I. Polat, Y. Feng, E. W. Moore, J. VanderPlas, D. Laxalde, J. Perktold, R. Cimrman, I. Henriksen, E. A. Quintero, C. R. Harris, A. M. Archibald, A. H. Ribeiro, F. Pedregosa, P. van Mulbregt, and SciPy 1.0 Contributors, ``SciPy 1.0: fundamental algorithms for scientific computing in Python'', Nat. Methods \textbf{17}, 261-272 (2020).  

\bibitem{Rummler2002} H. Rummler, ``On the distribution of rotation angles: How great is the mean rotation angle of a random rotation?'', Math. Intell. \textbf{24}, 6-11 (2002).  

\bibitem{BezansonEdelman2017} J. Bezanson, A. Edelman, S. Karpinski, and V. B. Shah, ``Julia: A fresh approach to numerical computing'', SIAM Rev. 59, 65 (2017).  

\bibitem{BenchmarkTools} \url{https://github.com/JuliaCI/BenchmarkTools.jl}

\bibitem{PythonCode} See \url{https://github.com/scipy/scipy/blob/v1.4.1/scipy/stats/_multivariate.py} for the original code.  

\bibitem{MarrinkRisselada2007}  S. J. Marrink, H. J. Risselada, S. Yefimov, D. P. Tieleman, A. H. de Vries, ``The MARTINI force field: Coarse grained model for biomolecular simulations'', J. Phys. Chem. B \textbf{111}, 7812-7824 (2007).  


\end{thebibliography}
\end{document}